\documentclass[%
 aip,
 rsi,%
 amsmath,amssymb,
 reprint,%
]{revtex4-1}

\usepackage{graphicx}
\usepackage{dcolumn}
\usepackage{bm}

\begin{document}


\title{Light-by-Light Scattering in a Photon-Photon Collider}

\author{T.~Takahashi}
\email{tohru-takahashi@hiroshima-u.ac.jp}
\thanks{Corresponding author}
\affiliation{AdSM Hiroshima University, 1-3-1 Kagamiyama, Higashi Hiroshima, Hiroshima 739-8530, Japan}
\author{G.~An}
\affiliation{Institute of High Energy Physics, Chinese Academy of Sciences, Beijing 100049, China }
\author{Y.~Chen}
\affiliation{Institute of High Energy Physics, Chinese Academy of Sciences, Beijing 100049, China }
\author{W.~Chou}
\affiliation{Institute of High Energy Physics, Chinese Academy of Sciences, Beijing 100049, China }
\author{Y.~Huang}
\affiliation{Institute of High Energy Physics, Chinese Academy of Sciences, Beijing 100049, China }
\author{W.~Liu}
\affiliation{Institute of High Energy Physics, Chinese Academy of Sciences, Beijing 100049, China }
\author{W.~Lu}
\affiliation{Department of Engineering Physics, Tsinghua University, Beijing 100084, China}
\author{J.~Lv}
\affiliation{Institute of High Energy Physics, Chinese Academy of Sciences, Beijing 100049, China }
\author{G.~Pei}
\affiliation{Institute of High Energy Physics, Chinese Academy of Sciences, Beijing 100049, China }
\author{ S.~Pei}
\affiliation{Institute of High Energy Physics, Chinese Academy of Sciences, Beijing 100049, China }
\author{C.~P.~Shen}
\affiliation{School of Physics and Nuclear Energy Engineering, Beihang University, Beijing 100191, China}
\author{B.~Sun}
\affiliation{School of Physics and Nuclear Energy Engineering, Beihang University, Beijing 100191, China}
\author{C.~Zhang}
\affiliation{Institute of High Energy Physics, Chinese Academy of Sciences, Beijing 100049, China }
\author{C.~Zhang}
\affiliation{Key Laboratory of Beam Technology and Materials Modification of the Ministry of Education, and College of Nuclear Science and Technology, Beijing Normal University, Beijing 100875, China }

\date{\today}

\begin{abstract}
We studied the feasibility of observing light-by-light scattering in a photon-photon collider based on an existing accelerator complex and a commercially available laser system.
We investigated the statistical significance of the signal over the QED backgrounds through a Monte Carlo simulation with a detector model. 
The study showed that light-by-light scattering can be observed with a statistical significance of 8 to 10 sigma in a year of operation, depending on the operating conditions. 
\end{abstract}

\maketitle

\section{\label{sec:introduction}Introduction}
Quantum electrodynamics (QED) is one of the most successful theories that describe electromagnetic interaction 
and has been tested with great precision. 
Although QED is a well-validated theory that underwent many experimental verifications, not all of its predictions have been observed to date.
 One of such areas is the interaction between real photons. 
Although the interactions between photons have been tested with great precision, all or a part of the photons involved in these interactions have been virtual photons. 
For example, electron-positron pair creation caused by a photon impinged on the media, which is one of the most famous and well-tested processes in QED,
 is an interaction between a real photon and electric fields in the media, characterizing a virtual photon.
The only experimental observation of pair creation by real photons was obtained through nonlinear QED interaction, 
performed by the E144 experiment in SLAC \cite{E144}.
Another phenomenon of interest is light-by-light scattering. A higher-order perturbation of QED predicts an elastic scattering between two photons. 
This phenomenon has been known since the conception of QED and the cross-section was calculated approximately 50 years 
ago\cite{tollis1,tollis2}. 
\begin{figure}
\includegraphics[width=8.0cm]{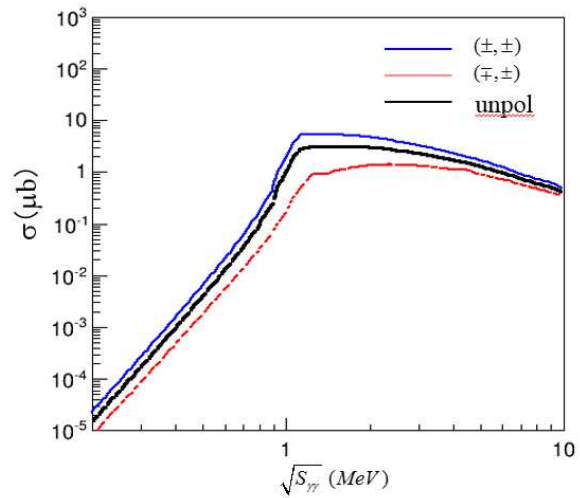}%
\caption{\label{fig:crosssection}
Cross-section of the process, $\gamma \gamma \to \gamma \gamma$, by the collision of circularly polarized photons
based on the formulas in \cite{tollis1,tollis2}; 
$(\pm,\pm)$  and $(\pm,\mp)$ stand for the combination of the helicities of the two colliding photons.}
\end{figure}
While several attempts to observe this phenomenon have been performed,  
no observation has been reported to date \cite{moulin,marklund}.
An experimental observation of this process in heavy-ion collisions at the LHC has been reported
by the ATLAS experiment in 2017 \cite{enterria,atlas}; 
however, the direct observation in collisions of real photons still remains to be reported. 

Recently, experiments to probe light-by-light scattering were proposed \cite{INFN,homma}. 
Both of them utilized laser-Compton scattering to generate photons at the center-of-mass energy range of approximately 1 MeV, 
where the cross-section of light-by-light scattering is at the maximum,
as shown in fig. \ref{fig:crosssection}. 
In ~\onlinecite{INFN}, the authors proposed an accelerator-based laser Compton facility; however, their proposal 
 has not been approved. 
In ~\onlinecite{homma}, the authors plan to use a laser-plasma accelerator to provide an electron beam for Compton scattering. 
It may be a good opportunity to perform this kind of experiments, 
but
 the electron beam facility with laser-plasma acceleration remains to be developed.
It must be pointed out that, in both proposals, the feasibility of observation of light-by-light scattering over 
possible backgrounds considering the detector response, has not been fully studied. 

In this article, we report on the feasibility of observation of light-by-light scattering through a Monte Carlo simulation study considering most of the issues, such as background processes, a detector model and design of a photon-photon collider based on the Beijing Electron Positron Collider (BEPC) accelerator complex in IHEP, China.

\section{\label{sec:section2} Photon-Photon Collider}

The concept of a photon-photon collider based on backward Compton scattering was first discussed in ~\onlinecite{gin1,gin2,gin3}.
Since then, it has been mainly discussed as an option for high energy electron-positron colliders
 \cite{NLCZDR,GLC,TESLA}.

\begin{figure}
\includegraphics[width=8.0cm]{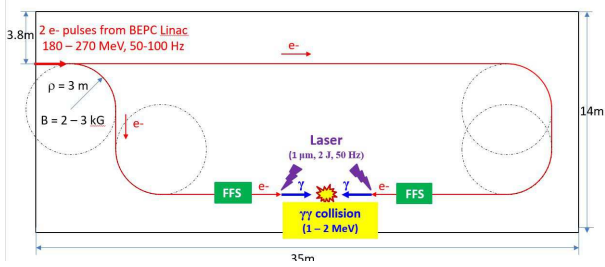}%
\caption{\label{fig:BEPC}
A planned layout of the $\gamma \gamma$ collider facility in the IHEP experimental hall 10}
\end{figure}

A planned layout of the experimental facility in an IHEP experimental hall is illustrated in fig. \ref{fig:BEPC}.
 The electron beams from the BEPC linac are introduced to the hall and divided into two arcs, 
and subsequently brought into a head-on collision at the interaction point (IP). 
The laser pulses are flashed onto the electron beams at the conversion points (CP) shortly before they cross the IP, in order to generate photon beams. 
The main parameters of the electron and laser beams are summarized in table \ref{tab:parameters}. 
\begin{table}
\caption{\label{tab:parameters}%
Laser and electron beam parameters for the proposed $\gamma \gamma$ collider.
The size and pulse length of the laser pulse were defined as the RMS of the intensity.
}
\begin{tabular}{llll}
\multicolumn{2}{l}{\textrm{Laser}}& 
\multicolumn{2}{l}{\textrm{Electron}}\\
Wave length ($\mu m$) & 1.054 &Energy(MeV) & 245 \\
Size at focus ($\mu m$) & 5 & Bunch charge (nC)  & 2(0.4)\\
Rayleigh Range ($\mu m$)  & 300 &  Size of IP ($\mu m$) & 2 \\
Pulse energy (J) & 2 (0.4) & emittance ($nm$)& $5.2$ \\
Pulse Length (ps) & 2 & beta* at IP (mm) & 727 \\
Repetition (Hz) & 100 & bunch length (mm) & 0.6 \\
Angle (mr) to e-beam & 167 & Repetition (Hz) & 100\\
IP-CP distance ($\mu m$) & 383 & crossing angle (mr) & 0 \\
Nonlinear parameter & \multicolumn{3}{l}{0.3(2 J)}\\
\end{tabular}
\end{table}

\begin{figure}
\includegraphics[width=9.0cm]{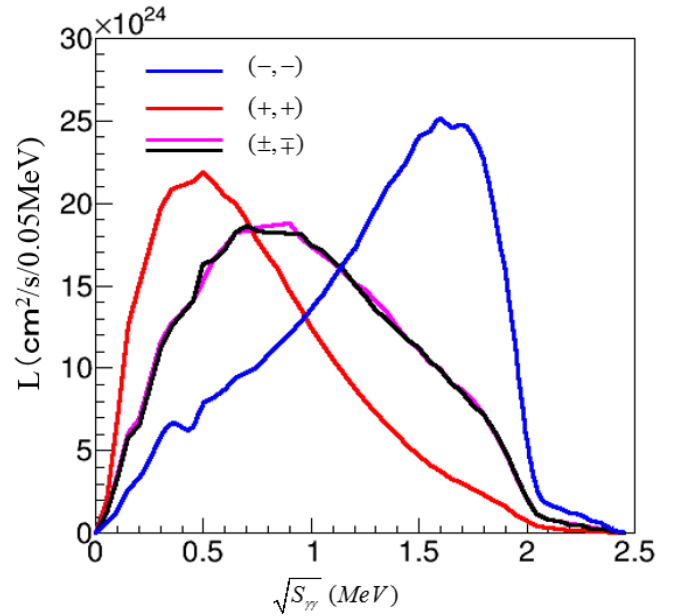}%
\caption{\label{fig:cainlum}
Differential luminosity calculated using CAIN with the parameters shown in table \ref{tab:parameters}.}
\end{figure}

We plan to use a photo-injector for the electron source to provide a low-emittance beam. 
The laser system is assumed to have a 0.4--2 J/pulse with a width of 2 ps in the RMS and 
a repetition rate of 100 Hz.
The differential luminosity calculated using CAIN 2.42 \cite{cain} is shown in fig. \ref{fig:cainlum}. 
The total luminosity of photon-photon collision is approximately $4.0 \times 10^{27}cm^{-2}s^{-1}$
for the laser pulse energy of 2 J and the electron bunch charge of 2 nC.

\subsection{\label{sec:ir}The Interaction region and detector}

\begin{figure}
\includegraphics[width=9.0cm]{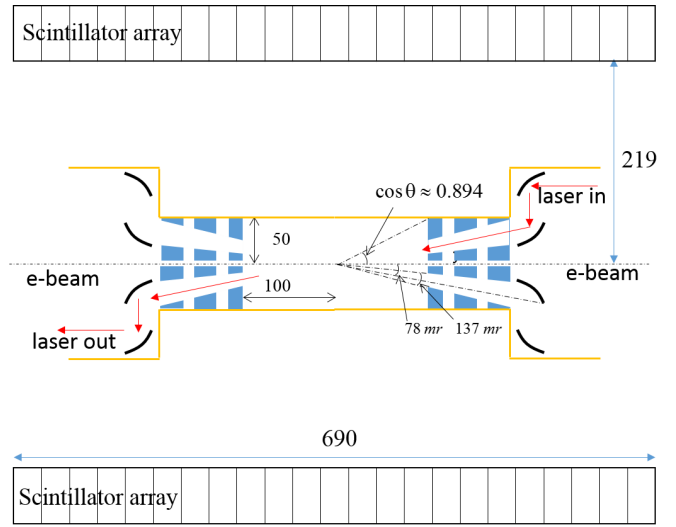}%
\caption{\label{fig:ir}
Cross-sectional view of the detector model.
The detector is a scintillator array made of CsI and plastic scintillator.
The final focus magnet of the radius of 50 mm is placed 100 mm from the interaction point. 
The laser pass (only one of the two) is indicated by arrows.}
\end{figure}

Fig. \ref{fig:ir} illustrates the interaction region of the detector. 
We assumed a final focus system which consists of three permanent magnets, similar to that discussed in the reference \cite{INFN}. The final focus magnet is placed 10 cm from the IP. The inner and the outer radius of the magnet are 3 mm and 50 mm, respectively. 
To introduce laser pulses to the electron beam, the magnets have holes of 137 mr in an angular aperture centered at 78 mr 
with respect to the beam axis and pointing to the IP. 
The angular aperture of the holes is approximately $\pm 4 \sigma$ of the angular divergence of the laser wave to ensure
good focusing property of the laser pulse at the CP. 
The beam pipe, made of 1-mm-thick beryllium, has an inner radius of 50 mm so that the final focus magnets are placed within the beam pipe.
 
The detector is a calorimetric system with a scintillator array. 
Each module of the array consists of a 2-mm-thick plastic scintillator backed by a 60-mm-thick CsI crystal of a trapezoidal shape. The cross-section of the innermost surfaces (plastic scintillator) of the module is 30 mm $\times$ 30 mm. The detector modules are arranged in a cylindrical shape along the beam axis. The number of modules is 23 in the beam direction, 46 in the plane perpendicular to the beam direction, and are 1058 in total (see fig. \ref{fig:ir}). 
The distance, $R$, of the front surface of the module from the IP is:
$$
R = \frac{w}{{2\tan \varphi/2 }} \approx 219.3mm
$$
 with $w$ and $\varphi $ being 30 mm and $2\pi /46$, respectively.
The detector model was implemented in the Geant4 toolkit to simulate the detector response for the events\cite{geant4}. After the simulation, the energy deposition in each detector module was smeared with a typical energy resolution (in RMS) of the CsI and plastic scintillator as:
$
\frac{{d\sigma }}{E} = \frac{{0.059}}{{\sqrt {E(MeV)} }}
$
for the plastic and
$
\frac{{d\sigma }}{E} = \frac{{0.024}}{{\sqrt {E(MeV)} }}
$
for the CsI scintillator, respectively.

\section{\label{eventgen} Event Rate and Monte Carlo Event Generation}
To generate Monte Carlo (MC) events for the simulation study, the event rate $N$ as a function of the polar angle $\theta$ was calculated as: 
\begin{equation}
\label{eq:evgen}
\begin{array}{l}
   \frac{{dN}}{{d\theta }}({\omega _{\gamma \gamma }},{a_{\gamma \gamma }},{h_{\gamma \gamma }})  \\ 
  =  \frac{{d{\sigma _{\gamma \gamma }}}}{{d\theta }}({\omega _{\gamma \gamma }},{h_{\gamma \gamma }})\frac{{dL}}{{d{\omega _{\gamma \gamma }}}}({\omega _{\gamma \gamma }},{a_{\gamma \gamma }},{h_{\gamma \gamma }})d{\omega _{\gamma \gamma }},
\end{array}
\end{equation}
which is a function of the center-of-mass energy of the colliding photon-photon system, $\omega_{\gamma \gamma}$, 
helicity combination of two photons, $h_{\gamma\gamma}$ 
and energy asymmetry of two photons, $a_{\gamma\gamma}$, 
where  
${h_{\gamma \gamma }} = ( + , + ),\;( + , - ),\;( - , + ),\;( - , - )$
denotes the helicity combination of the two colliding photons 
and 
 ${a_{\gamma \gamma }} \equiv \frac{{e_\gamma ^1 - e_\gamma ^2}}{{e_\gamma ^1 + e_\gamma ^1}}
$ 
is the energy asymmetry of the two colliding photons.
To perform the calculation, the differential luminosity
$
\frac{{dL}}{{d{\omega _{\gamma \gamma }}}}({\omega _{\gamma \gamma }},{a_{\gamma \gamma }},{h_{\gamma \gamma }})$
was estimated with CAIN in two-dimensional space in $\omega_{\gamma\gamma}$, $a_{\gamma \gamma }$
for each helicity combination, ${h_{\gamma \gamma }}$.
 The cross-section of the process $\gamma \gamma \to \gamma \gamma$, $\frac{{d{\sigma_{\gamma \gamma }}}}{{d\theta }}({\omega _{\gamma \gamma }},{h_{\gamma \gamma }})$, 
was calculated according to the formula given in the reference \cite{tollis1,tollis2}
in the center-of-mass system and boosted to the laboratory system with the energy asymmetry $a_{\gamma \gamma}$.
Subsequently, the Monte Carlo events were generated in a five-dimensional phase space in $ \theta ,\varphi ,{\omega _{\gamma \gamma }},{a_{\gamma \gamma }},{h_{\gamma \gamma }}$.
As the background process for $\gamma \gamma \to \gamma \gamma$, we considered following processes:
\begin{itemize}
\item $\gamma \gamma  \to {e^ + }{e^ - }$ Breit-Wheeler process
\item $\gamma \gamma  \to {e^ + }{e^ - }\gamma$ 
and $\gamma \gamma  \to {e^ + }{e^ - }\gamma \gamma $ 
Breit-Wheeler process with the final state radiations 
\item ${e^ - } \gamma  \to {e^ - } \gamma $ Compton scattering of a Compton photon and a beam electron
\item ${e^ - } \gamma  \to {e^ + }{e^ - }{e^ - }$ trident process with a Compton photon and a beam electron
\item $ {e^ - }{e^ - }  \to {e^ - }{e^ - }$ Moller scattering of beam electrons.
\end{itemize}
We adopted WHIZARD\cite{whizard} for the cross-section calculations and subsequent event generations for background processes. 
The photon-photon, photon-electron, and electron-electron luminosity distributions were implemented in 
WHIZARD via its built-in utility program CIRCE2.
The effective production cross-section was defined as:
 $$
{\sigma _{eff}} \equiv \frac{1}{{L_{ij}^{tot}}}\sum\limits_{{h_{ij}}} {\int {\sigma ({\omega _{ij}},{h_{ij}})} \frac{{d{L_{ij}}}}{{d{\omega _{ij}}}}({\omega _{ij}},{h_{ij}})d{\omega _{ij}}} .
$$
In the above expression, the total luminosity is calculated using the results of CAIN as:
$$
L_{ij}^{tot} \equiv \sum\limits_{{h_{ij}}} {\int {\frac{{d{L_{ij}}}}{{d{\omega _{ij}}}}({\omega _{ij}},{h_{ij}})d{\omega _{ij}}} } ,
$$
where $ij$ stands for the initial beams, i.e, Compton photon or the electron beam.
All cross-sections except for $\gamma \gamma  \to \gamma \gamma$ were calculated with a kinematic cut, such that 
at least one final state particle must be within the angular acceptance of the detector($\left| {\cos \theta } \right| < 0.8944$)
(see fig. \ref{fig:ir}). 
The calculated effective cross-section for each process is:
\begin{equation}
\begin{array}{l}
   {\sigma _{eff}}(\gamma \gamma  \to \gamma \gamma )  = 1.5\mu b  \\
  {\sigma _{eff}}(\gamma \gamma  \to {e^ + }{e^ - })  = 5.7 \times {10^4}\mu b  \\
  {\sigma _{eff}}(\gamma \gamma  \to {e^ + }{e^ - }\gamma )  = 4.7 \times {10^3}\mu b  \\ 
  {\sigma _{eff}}(\gamma \gamma  \to {e^ + }{e^ - }\gamma \gamma )  = 2.1 \times {10^2}\mu b  \\ 
  {\sigma _{eff}}({e^ - }\gamma  \to {e^ - } \gamma)  = 8.9 \times {10^2}\mu b  \\
  {\sigma _{eff}}({e^ - }\gamma  \to {e^ + }{e^ - }{e^ - })  = 3.7 \times {10^3}\mu b  \\
  {\sigma _{eff}}({e^ - }{e^ - } \to {e^ - }{e^ - })  = 2.1 \times {10^1}\mu b
\end{array}
\end{equation}

\section{\label{sec:pileups}Event Pileups}
In order to ensure proper detector operation and subsequent data analysis, we must suppress the event pileups in 
a bunch collision below a reasonable level.  
In general, the acceptable rate of the pileups depends on the details of data analysis
 and the probability of the pileups and total (integrated) luminosity depend on the laser and electron beam parameters.

Therefore, in this analysis, we assumed that the average number of events per bunch for 
$\gamma \gamma  \to {e^ + }{e^ - }$,which has the largest cross-section among the background processes, is
$N_b(\gamma \gamma  \to {e^ + }{e^ - })=0.1$. 
The corresponding bunch luminosity for $\gamma \gamma$ collision, $L_b^{\gamma \gamma}$ is,
$$
L_b^{\gamma \gamma} = 0.1/{\sigma _{eff}}(\gamma \gamma  \to {e^ + }{e^ - }) \approx  1.76 \times 10^{24} cm^{-2}
$$ 
According to the luminosity calculation with CAIN, 
the bunch luminosity for the laser pulse of 2 J/pulse and 
the electron bunch charge of 2 nC/bunch (see table\ref{tab:parameters}) is 
 $\approx 4 \times 10^{25} cm^{-2}$. 
Thus, we must decrease $L_b^{\gamma \gamma}$ by a factor of about 23.
We assumed to realize this luminosity either by reducing the electron charge 
to 0.4 nC/bunch with the laser energy of 2 J/pulse, or by reducing the laser energy to 0.4 J/pulse 
with the electron charge of 2 nC/bunch.
Hereafter, we refer to the former case as (0.4 nC, 2 J) and to the later as (2 nC, 0.4 J).  
When estimating the number of events for two cases, we assumed following conditions
\begin{itemize}
\item When changing the electron bunch charge, the ratios of the number of events for the signal and backgrounds are the same as 
the estimation with 2nC/bunch and 2 J/pulse case, since bunch luminosities of all combinations, $L_b^{\gamma \gamma}$, $L_b^{e^- \gamma}$, $L_b^{e^- e^-}$, have the same charge dependence, i.e,
approximately proportional to the square of the electron bunch charge. 
\item When changing the laser pulse energy, $L_b^{\gamma \gamma}$ is approximately  proportional to the square of the laser pulse energy,
$L_b^{e^- \gamma}$ is linearly proportional to the laser pulse energy and $L_b^{e^- e^-}$ is independent of the laser pulse energy.
The number of events were estimated accordingly.
\end{itemize}
   Consequently, the expected number of events per second at 100 Hz beam repetition rate, $N_{sec}$, is
\begin{equation}
\begin{array}{l}
   N_{sec}(\gamma \gamma  \to \gamma \gamma )  = 2.7 \times 10^{-4} \\
   N_{sec}(\gamma \gamma  \to {e^ + }{e^ - })  = 10  \\
   N_{sec}(\gamma \gamma  \to {e^ + }{e^ - }\gamma )  = 8.3 \times 10^{-1}  \\ 
   N_{sec}(\gamma \gamma  \to {e^ + }{e^ - }\gamma \gamma )  = 3.7 \times 10^{-2}\\ 
   N_{sec}({e^-} \gamma  \to {e^ - } \gamma)  = 0.3 (1.5) \\
   N_{sec}({e^-} \gamma  \to {e^ + }{e^ - }{e^ - })  = 1.1(5.5) \\
   N_{sec}({e^ - }{e^ - } \to {e^ - }{e^ - })  = 2.5 \times 10^{-3} (6.3 \times 10^{-2})
\end{array}
\end{equation}
In the above expressions, the numbers in parentheses are for (2 nC, 0.4 J) and the others are 
for (0.4 nC, 2 J), i.e., the number of events for $e^- \gamma$, and $e^-  e^-$ collisions with (2 nC, 0.4 J), 
are 5 times and 25 times larger than those with  (0.4 nC, 2 J), respectively.
Summing up the contribution of other processes, 
the total number background events per bunch, $n_{pl}$, is expected to be approximately 0.12 with (0.4 nC, 2 J)
and 0.18 with (2n C, 0.4 J), respectively.
The pileups are not implemented in the current detector simulation. 
We, instead, reduced the total number of events by $e^{-n_{pl}}$, 
assuming all the pileup events are rejected 
by the clustering analysis described in the next section.

\section{\label{sec:analysis}Data Analysis}
\subsection{\label{sec:overview} Analysis overview}
\begin{figure}
\includegraphics[width=9.0cm]{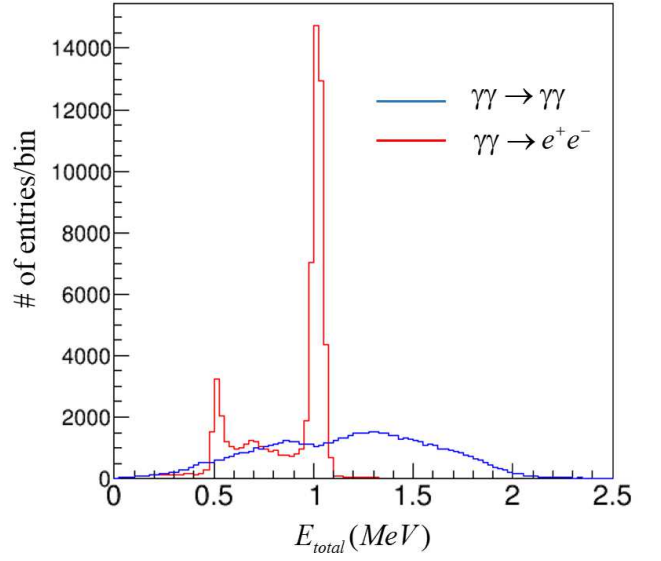}%
\caption{\label{fig:edepall} Total energy deposition observed by the detector. Clear peaks caused by a positron annihilation in the materials for
  $\gamma \gamma  \to  {e^ + }{e^ - }$ process are observed; the peak around 1 MeV represents the case where two 0.511 MeV photons are destructed, 
while the peak around 0.511 MeV is indicative of the case in which one of the two photon escapes from the detector.
}
\end{figure}

In fig. \ref{fig:edepall}, we show the total energy deposition in the detector for 
$\gamma \gamma  \to \gamma \gamma$ and $\gamma \gamma  \to  {e^ + }{e^ - }$ processes
after the detector simulation described in section \ref{sec:ir}.

\begin{figure}
\includegraphics[width=7.0cm]{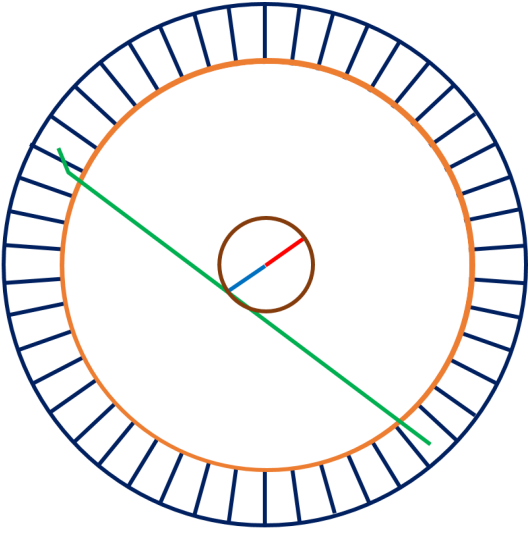}%
\caption{\label{fig:display}Illustration of a typical  $\gamma \gamma  \to  {e^ + }{e^ - }$ event.
A positron is absorbed in the beam pipe and generates back to back photons.}
\end{figure}

The peaks around 0.5 MeV and 1 MeV are observed for $\gamma \gamma \to  {e^ + }{e^ - }$ events.
These peaks are attributed to the annihilation of a positron in materials. 
As the center-of-mass energy of the $\gamma \gamma$ collision is near the threshold of pair creation, the momenta of the electron and positron are of the order of a few hundred keV at most.
Therefore, virtually, all electrons and positrons stop inside the material around the IP, even in the beam pipes of 1-mm-thick beryllium. 
When a positron stops in materials, it generates back-to-back photons of approximately 0.511 MeV. 
This phenomenon resembles the signal events and its discrimination is crucial for the signal detection.
It should be noted that, because the event signatures of both processes are back-to-back photons, identification of the charged 
particle will not help improve the situation.
A typical $\gamma \gamma  \to  {e^ + }{e^ - }$ event observed in the detector is shown in fig. \ref{fig:display}.
Two photons are 
generated in back to back direction from a point in the beam pipe, where a positron is absorbed, 
while the electron absorbed in the beam pipe had no effect. 
In order to discriminate the background photons from the beam pipe, and the signal from the IP, we increase the radius of the beam pipe to 50 mm so that these two processes can be discriminated from one another by 
defining proper observables, as described in section \ref{sec:evanalysis}. 

\subsection{\label{sec:evanalysis}Event analysis}
We generated the MC events for signal and background processes. 
The number of event samples were, 
2,100 k for $\gamma \gamma  \to  {e^ + }{e^ - }$,
1,100 k for $\gamma \gamma  \to  {e^ + }{e^ - } \gamma$ and 
 ${e^ - } \gamma  \to  {e^ + }{e^ - }{e^ - }$, 
and 200 k for the remaining signal and background processes.
After simulating the detector response by Geant4, momentum vectors, $ {\vec p_i}$, were defined for each hit 
in a detector module
in an event as: 
$$
{\vec p_i} = {E_i}\,{\vec r_i}
$$
where $E_i$ is the energy deposition in $i_{th}$ plastic or CsI module and ${\vec r_i}$ is the position of 
the innermost surface of the module. 
With these vectors, two jets,  ${\vec P_{1(2)}}$, which represent the final state particles for an event, 
were reconstructed by applying the forced two-jet clustering technique with the Durham algorithm \cite{durham,fastjet}.
(${\vec P_{1}}$ and ${\vec P_{2}}$ are chosen such that ${\vec P_{1}}$ is the jet of the higher energy.)
For further analysis, the following observables are defined using ${\vec P_{j}}$ as: 
\begin{enumerate}
\item ${E_j} \equiv \left| {{{\vec P}_j}} \right|$; Energy of each jet; 
\item $E = \sum {{E_j}} $; Total energy deposition of an event; 
\item $\cos \,{\theta _{{P_1}}}$
\item $\cos \,{\theta _{{P_2}}}$
\item $\,{\theta _{acl}} \equiv \pi  - {\vec P_1}\angle {\vec P_2}$; acollinearity angle
\item $\,{\theta _{acp}} \equiv \pi  - P_1^ \bot \angle P_2^ \bot $ ; acoplanarity angle 
\item  $E_1^{psc}$;  Energy deposition on the plastic scintillator of  ${\vec P_{1}}$ 
\item  $E_2^{psc}$;  Energy deposition on the plastic scintillator of  ${\vec P_{2}}$ 
\end{enumerate}
In the above expression, ${\vec P_1}\angle {\vec P_2} $ stands for the opening angle of two vectors,  ${\vec P_1}$, ${\vec P_2}$.

\begin{table*}
\caption{\label{tab:results}
Expected number of signals and backgrounds with one year ($10^7$ s) of operation 
at the maximum statistical significance (BDT output = 0.295). 
The numbers in the first row are for the case with (electron bunch charge, laser pulse energy) = (0.4 nC, 2 J)
and those in the second row are for (2 nC, 0.4 J). 
(See section \ref{sec:pileups}  for details.)
Errors are from statistics of the MC events.
}
\begin{tabular*}{\textwidth}{@{\extracolsep{\fill}}llllllll@{}}
$\gamma \gamma \to \gamma \gamma$ &
$\gamma \gamma  \to {e^ + }{e^ - }$ &
$\gamma \gamma  \to {e^ + }{e^ - } \gamma$ &
$\gamma \gamma  \to {e^ + }{e^ - } \gamma \gamma$ &
${e^ - } \gamma  \to {e^ - }\gamma$ &
${e^ - } \gamma  \to {e^ + }{e^ - }{e^ - }  $ &
${e^ - } {e^ - }  \to {e^ - }{e^ - }  $ &
bg total \\
\hline \\
383.1 $\pm$ 2.8 &
530 $\pm$150 &
15 $\pm$10 &
6.5$\pm$4.6 &
 $<$ 32 &
131 $\pm$ 36 &
$< $ 0.25 &
680 $\pm$ 160 \\
364.0 $\pm$ 2.6 &
510 $\pm$150 &
14 $\pm$10 &
6.2$\pm$4.4 &
 $<$ 160 &
620 $\pm$ 170 &
$< $ 6 &
1150 $\pm$ 270 \\
\hline \\
\end{tabular*}
\end{table*}

\begin{figure}
\includegraphics[width=8.5cm]{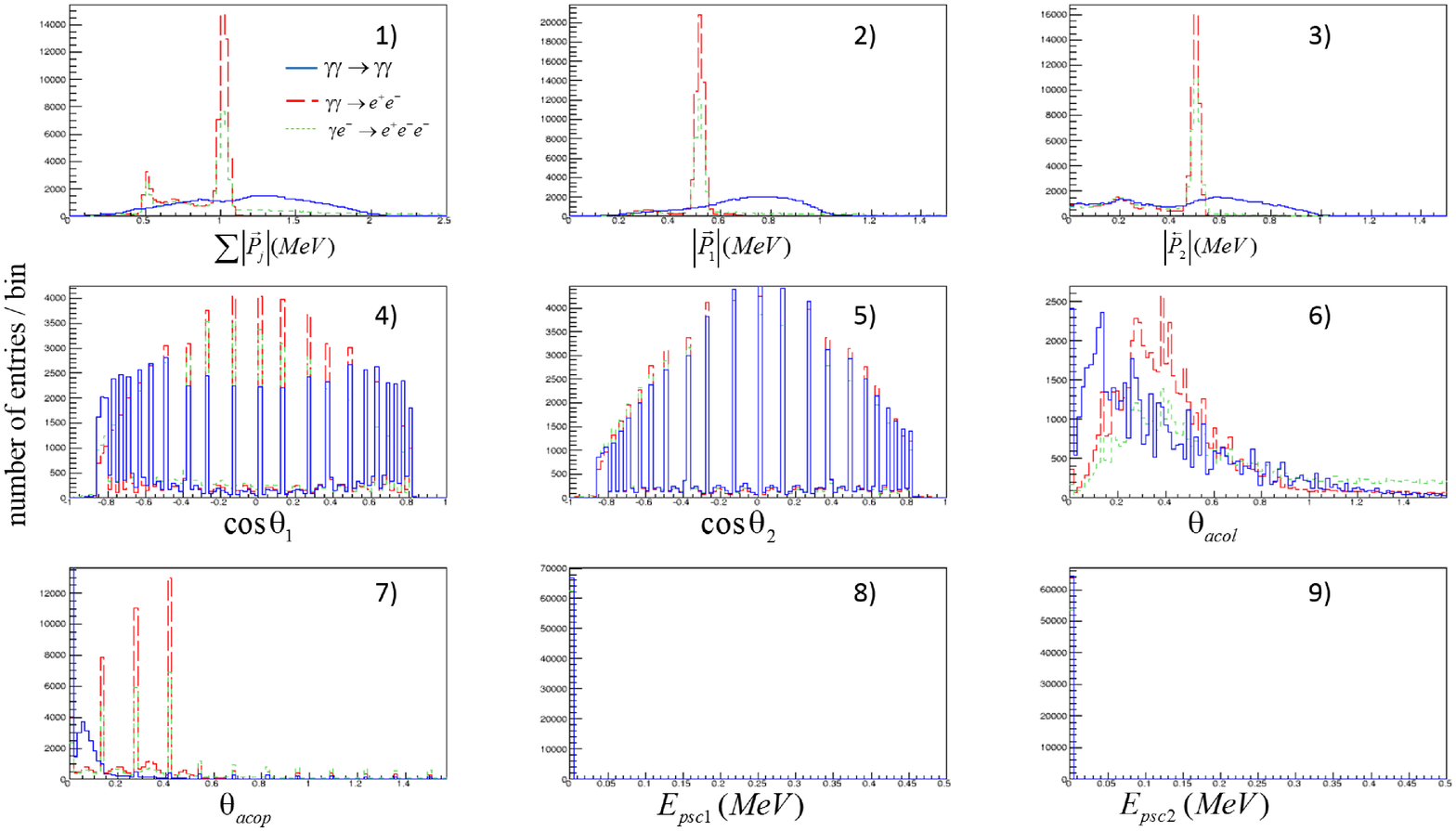}%
\caption{\label{fig:distribution}Distributions of observables used for the analysis, signal (blue-solid), 
$\gamma \gamma \to e^+,e^- $ (red, dashed), and $\gamma e^- \to e^+ e^- e^- $ (green-dots).
For visualization purposes, the number of entries in each figure is a result of 100k simulated events and does not represent the expected number of events in the experiment.
The numbers on each panel correspond to definitions of valuables
in the text.}
\end{figure}

The distribution of each observable for the signal and background events are shown in fig. \ref{fig:distribution}. 

In order to discriminate signal from backgrounds, 
we first applied the following criteria:
\begin{itemize}
\item  $E  <  2.5$  GeV
\item $E_1^{psc}=E_2^{psc}=0$
\item $\,{\theta _{acp}} < 0.15 $
\end{itemize}
The requirement for the total energy deposition is effective to discriminate $e^- e^- \to e^- e^-$   and
the cuts for the energy deposition in the plastic scintillators are essential to reduce contributions from 
charge tracks from the ${e^ - } \gamma  \to {e^ + }{e^ - }{e^ - }  $ process.
The acoplanarity angle cut is applied to improve efficiency of the selection optimization.

\begin{figure}
\includegraphics[width=8.5cm]{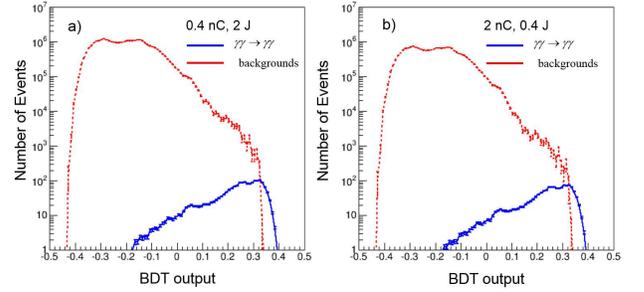}%
\caption{\label{fig:BDT} Expected BDT distribution for $\gamma \gamma \to \gamma \gamma$ (blue-solid) and 
sum of all background events (red-dashed) with one-year operation
for the operating conditions of (0.4 nC, 2 J) a), and (2 nC, 2 J) b).
The errors on data points are calculated with the statistics of MC events.}
\end{figure}

\begin{figure}
\includegraphics[width=8.0cm]{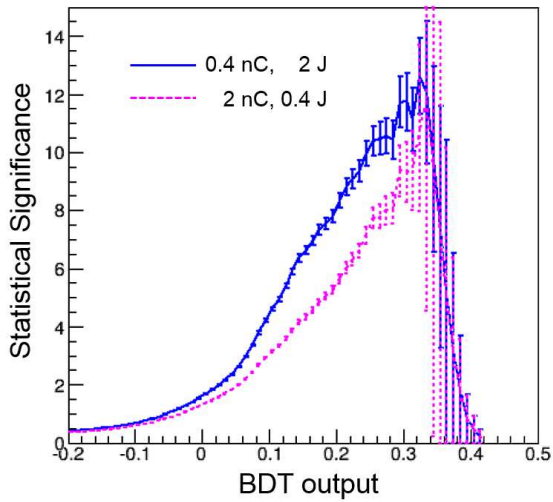}%
\caption{\label{fig:result} Expected statistical significance of $\gamma \gamma \to \gamma \gamma$ 
signal over the background after one year of operation for the operating conditions of 
(0.4 nC, 2 J) (blue-solid), and (2 nC, 0.4 J) (purple-dashed).}
\end{figure}

  To optimize event selection, we used the Boosted Decision Tree (BDT) implemented 
in the Toolkit for Multi-Variable Analysis (TMVA) in ROOT\cite{root}. 
For each signal and background event, 100k events were used to train the BDT and remaining events were used to test the performance of the event selection. 
The outputs from the BDT analysis are shown in fig. \ref{fig:BDT}.
Using the number of events survived for the signal and the background, 
the expected statistical significance, $sig$ is defined as:
$$
sig \equiv \frac{{{N_S}}}{{\sqrt {{N_S} + {N_B}} }}
$$
The significance was calculated with the expected number of events for a one-year operation ($10^7s$) of the experiment. The calculated significance is plotted as a function of the BDT output in fig.\ref{fig:result}. 
The error in the plots is the statistical uncertainty of the MC events. 
We obtained the best significance at the BDT output of 0.295, 
and the number of expected events for the signal and background are summarized in table \ref{tab:results}.
From the number of events, the statistical significance is estimated to be
\begin{equation}
sig = \left\{ {\begin{array}{*{20}{c}}
{11.72 \pm 0.88\;\quad {\rm{(0}}{\rm{.4nC,}}\;{\rm{2J)}}}\\
{\;\;9.35 \pm 0.86\;\quad {\rm{(2nC,}}\;{\rm{0}}{\rm{.4J)}}}
\end{array}} \right.
\end{equation}
The error is estimated from the statistics of MC events (see table \ref{tab:results}).
As a result, the expected statistical significance for observing $\gamma \gamma \to \gamma \gamma$ 
over the background processes is:
\begin{equation}
\begin{array}{c}
10.0 < sig < 13.4\;\quad {\rm{(0}}{\rm{.4nC,}}\;{\rm{2J)}}\\
\;\,7.7 < sig < 11.0\;\quad {\rm{(2nC,}}\;{\rm{0}}{\rm{.4J)}}
\end{array}
\end{equation}
with the 95 \% confidence interval,
or, roughly,  
we expect about 8 to 10 sigma excess of the signal over the background with the probability of 
0.975 after one-year ($10^7$ s) operation.

\section{Conclusion}
We demonstrated the feasibility of observing light-by-light scattering in a photon-photon collider based on a design with existing facility and feasible technology. 
An expected statistical significance to observe the signal has been investigated for the first time with a realistic luminosity distribution estimated using an accelerator lattice and laser optics, a detector model that can be constructed using reasonable resources, and a Monte Carlo data analysis considering most of the possible background processes.
The results showed that light-by-light scattering can be observed with a statistical significance of 8 to 10 sigma after one year of operation depending on the operating conditions, which is well above the level of discovery.

In this study, we have shown that light-by-light scattering can, in principle, be measured with a 
gamma-gamma collider at the CMS energy around 1 MeV.
We can, however, consider further developments and optimizations toward the construction of
the experimental facility.
Methods to manage pileup events must be studied in detail. 
The number of pileup events depends both on the laser pulse energy and the electron bunch charge, while  
the acceptable rate of the pileups must be investigated quantitatively by detector simulation. 
A comprehensive study closely related with hardware design is necessary.
The measurement of luminosity is another issue. 
It is necessary to measure the luminosities not only for $\gamma \gamma$ collision but also
for $e^{-} \gamma$ and $e^- e^-$ collisions. 
In the analysis we report in this paper, we trained the BDT to maximize $\gamma \gamma \to \gamma \gamma$ 
signal over the background. 
It is straightforward to train the BDT to enhance other processes such as 
  $\gamma \gamma \to e^+ e^-$ or $e^- \gamma \to e^- \gamma$ or  
$e^- \gamma \to e^- e^- e^+$ or $e^-  e^- \to e^- e^-$ to monitor the luminosity.  
It is also possible to turn on/off the laser pulses to measure each collision separately. 
As for the detector system, further R\&D, including the machine detector interface, laser optics, 
readout electronics and data acquisition system, is necessary.
The necessity of a tacking detector and/or a solenoid magnet are important issues to be discussed, 
since they could significantly improve the detector performance. 
 
In addition to the observation of light-by-light scattering, we note that
the Breit-Wheeler process has not been observed by the collision of real photons.
Furthermore, the proposed photon-photon collision system can be realized as a facility to observe non-linear effect in QED, 
such as a shift in the high energy peak and/or multiple photon absorption in Compton scattering.

\smallskip
\begin{acknowledgements}
This work is supported by National Natural Science Foundation of China (11655003), 
Innovation Project of IHEP (542017IHEPZZBS11820),  
and in part by the CAS Center for Excellence in Particle Physics (CCEPP).
\end{acknowledgements}

\end{document}